\documentclass[letter]{aa} 
\usepackage{graphicx}
\usepackage{txfonts}
\usepackage[]{natbib}

\begin{document}
   \title{Lepton models for TeV emission from SNR RX J1713.7-3946}


   \author{
    Zhonghui Fan\inst{1}
    \and
    Siming Liu\inst{2}
    \and
    Qiang Yuan\inst{3}
    \and
   Lyndsay Fletcher\inst{2}}

   \institute{Department of Physics, Yunnan University, Kunming
   650091, Yunnan China; fanzh@ynu.edu.cn
   \and
   Department of Physics and Astronomy, University of
   Glasgow, Glasgow G12 8QQ, UK; sliu@astro.gla.ac.uk
   \and
    Key Laboratory of Particle Astrophysics, Institute of
    High Energy Physics, Chinese Academy of Sciences, Beijing 100049, China}

\authorrunning{Z. H. Fan et al.}
\titlerunning{Lepton models for SNR RX J1713.7-3946}

\date{Received 8 June 2010 / Accepted 5 July 2010}


  \abstract
{}
{SNR RX J1713.7-3946 is perhaps one of the best observed shell-type supernova remnants
with emissions dominated by energetic particles accelerated near the shock front. The
nature of the TeV emission, however, is an issue still open to investigation.}
{We carried out a systematic study of four lepton models for the TeV emission with the
Markov chain Monte Carlo method.}
{It is shown that current data already give good constraints on the model parameters.
Two commonly used parametric models do not appear to fit the observed radio, X-ray,
and $\gamma$-ray spectra. Models motivated by diffusive shock acceleration and by
stochastic acceleration by compressive waves in the shock downstream give comparably
good fits. The former has a sharper spectral cutoff in the hard X-ray band than the
latter. Future observations with the {{\it HXMT}} and {{\it NuSTAR}} may distinguish these
two models.}
   {}


   \maketitle
%

\section{Introduction}

Energetic particles produce radiations over a broad energy range
under typical astrophysical circumstances. To study the energetic
particle population in high-energy astrophysical sources,
observations over a broad spectral range are needed.
Multi-wavelength observations of individual sources
usually give sparse spectral data points, which can be fitted
reasonably well with a simple power-law or broken power-law
function. A featureless power-law distribution implies the lack of
characteristic scales or distinct processes in the system. The
nature of particle acceleration processes are still a matter of
debate nearly a century after the discovery of high-energy particles
from the outer space \citep{b09}.

Recent progress in observations of high-energy astrophysical
sources has resulted in more detailed spectral coverage, and
spectral features start to emerge. These features are produced by
the underlying physical processes with well-defined characteristic
scales in space, time, and/or energy, and they play essential roles in
advancing our understanding of the acceleration mechanism
\citep{b88}. Diffusive astrophysical shocks are considered as one of
the most important particle accelerators \citep{t80, d83, b96}. The
acceleration of particles by compressive motions of astrophysical
flows is also a very generic process \citep{p88}. Both processes may
be important in the acceleration of particles dominating the
radiative characteristics of the shock associated with SNR RX
J1713.7-394 \citep{f10}. Detailed X-ray and $\gamma$-ray
observations of this source have revealed several features: 1) the
X-ray and $\gamma$-ray emissions are well correlated in space
\citep{a06, p08}; 2) the emission spectrum shows a clear high-energy
cutoff in both the X-ray and TeV bands \citep{a07}, and the cutoff in
hard X-rays appears to be sharp \citep{t08}, implying a sharp cutoff
in the electron distribution producing the observed radio to X-ray
spectrum through the synchrotron process; 3) the $\gamma$-ray
spectrum has a broad convex shape \citep{f09}; 4) there are bright
X-ray filaments with a width of $\sim0.1$ light year varying on a
timescale of about a year \citep{u07}. \citet{l08} showed that these
emissions may be attributed to a single population of electrons with
the $\gamma$-rays produced via the inverse Compton scattering of the
background photons by relativistic electrons.

In light of recent high-resolution observations with {\it Suzaku} and {\it Fermi},
we use the Markov chain Monte Carlo (MCMC) method to constrain parameters of
four possible lepton models. It is shown that simple models with a
power-law electron distribution, which cuts off at TeV energies, give a
poor fit to either the X-ray or $\gamma$-ray spectrum. Motivated by the
mechanism of diffusive shock acceleration (DSA) and stochastic acceleration
(SA) by compressive motions, we introduce two alternative models, both of
which give acceptable fits to the observed spectra (Section \ref{models}).
In Section \ref{dis}, we discuss the implications of these results and
future work needed to improve these models. Conclusions are drawn in Section \ref{con}.

\section{Spectral fits with lepton models}
\label{models}

In the lepton scenario, the radio to X-ray emissions are produced through the
synchrotron process of relativistic electrons, and the $\gamma$-ray emission
is produced through the inverse Compton scattering of the background radiation
fields by the same electron population. If the background magnetic fields are
relatively uniform, similar to the background radiation fields, the model can
naturally explain the spatial correlation between the X-rays and $\gamma$-rays
\citep{p08}. As pointed out by \citet{l08}, the magnetic field required to
produce the X-ray flux agrees with what is required to produce an X-ray
spectral cutoff by an electron population cutting off in the TeV energy range
\citep{a07, e10}.

There are at least four parameters in a specific model. In this work we employ
an MCMC technique suitable for multi-parameter determination to
search all the model parameters. The
Metropolis-Hastings sampling algorithm is adopted when determining the jump
probability from one point to the next in the parameter space \citep{m03}.
The MCMC approach, which is based on the Bayesian statistics, is superior to
the grid approach with a more efficient sampling of the parameter space of
interest, especially for  high dimensions. The algorithm ensures that the
probability density functions (PDF) of model parameters can be asymptotically
approached with the number density of samples.
Starting with an initial parameter set ${\bf P_0}$, which should
lie in a reasonable parameter space, one obtains the likelihood function of
${\bf P_0}$: $\mathcal{L}_0({\bf P_0})\propto \exp
\left(-{\chi^2({\bf P_0})}/{2}\right)$. Then another parameter set
${\bf P_1}$ is randomly selected with the corresponding likelihood function
$\mathcal{L}_1$. This parameter set is accepted with the probability
$\alpha=\min\{1,{\mathcal{L}_1}/{\mathcal{L}_0}\}$. If the new point is
accepted, it becomes the starting point for the next step, otherwise, we reset
${\bf P_1}={\bf P_0}$. This procedure is repeated to derive a parameter
chain that determines the PDF of model parameters. More details of this
procedure can be found in Neil (1993), Gamerman (1997), and MacKay (2003).

\begin{figure}
\centering
\includegraphics[width=9 cm]{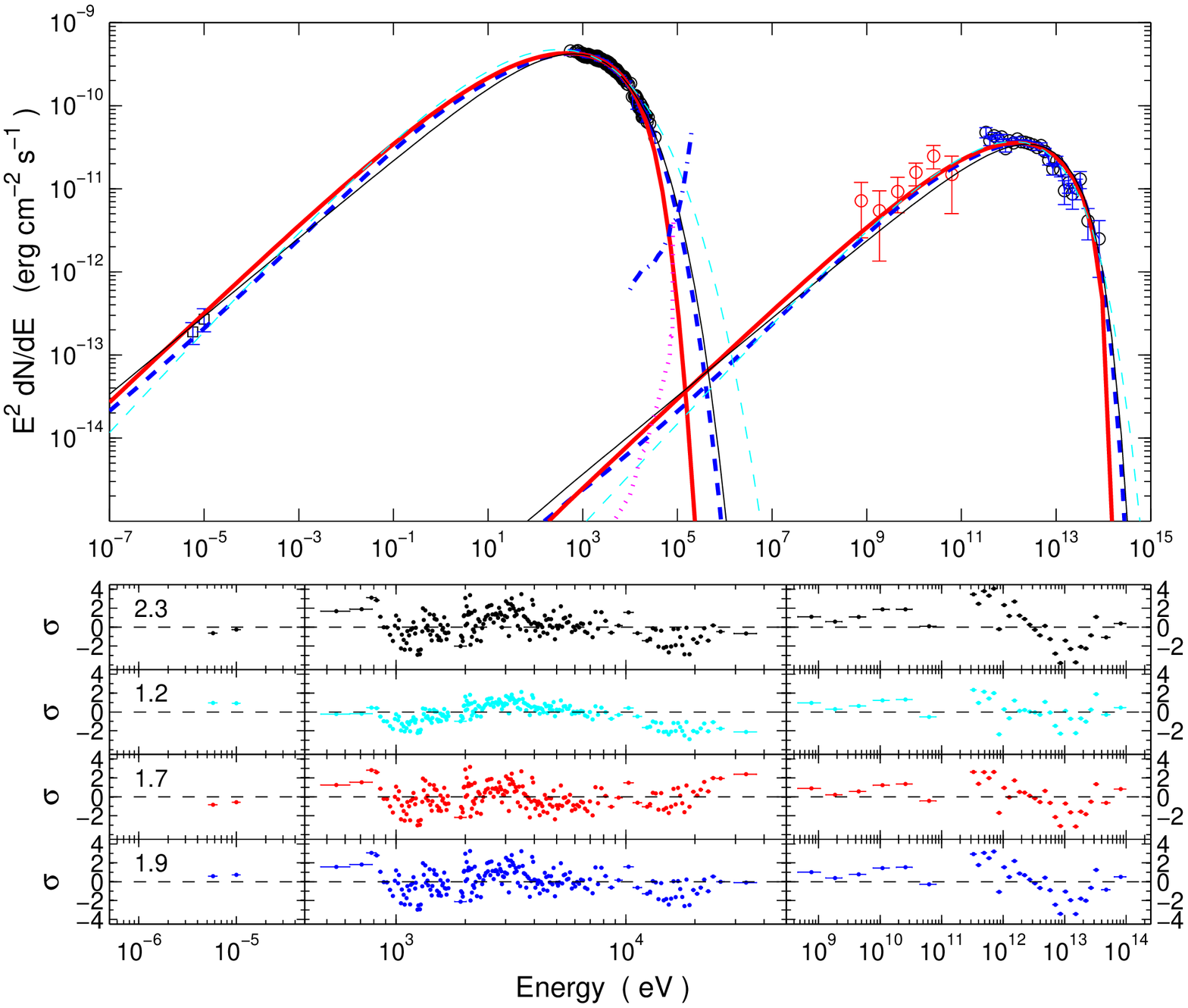}
\caption{Model fit to the spectrum of SNR RX J1713.7-3946. The TeV data is from
\citet{a07}, the {\it Fermi} data is from \citet{f09}, and the X-ray
data is from \citet{t08} (Thanks to Jun Fang). The thick solid, dashed
line and the thin solid, dashed line in the upper panel correspond to the
diffusive shock, stochastic acceleration model, and the model with an exponential,
more gradual cutoff in the electron distribution, respectively. The low and
high-energy spectral hump are produced by relativistic electrons through the
synchrotron and inverse Comptonization process, respectively. The lower panels
show the normalized residuals. From top to bottom, they are for the model with
an exponential and more gradual cutoff, the DSA, and SA model. For the model with
a gradual cutoff, the X-ray errors have been artificially increased by a factor
of 2 to improve the fit.  The reduced $\chi^2$ of these fits are indicated in
the top-left corner. See the text for details. The instrumental sensitivity of
{\it NuSTAR} and {\it HXMT} are indicated by the dotted and dot-dashed lines,
respectively.
\label{spec}}
\end{figure}

\begin{figure}
\centering
\includegraphics[width=9 cm]{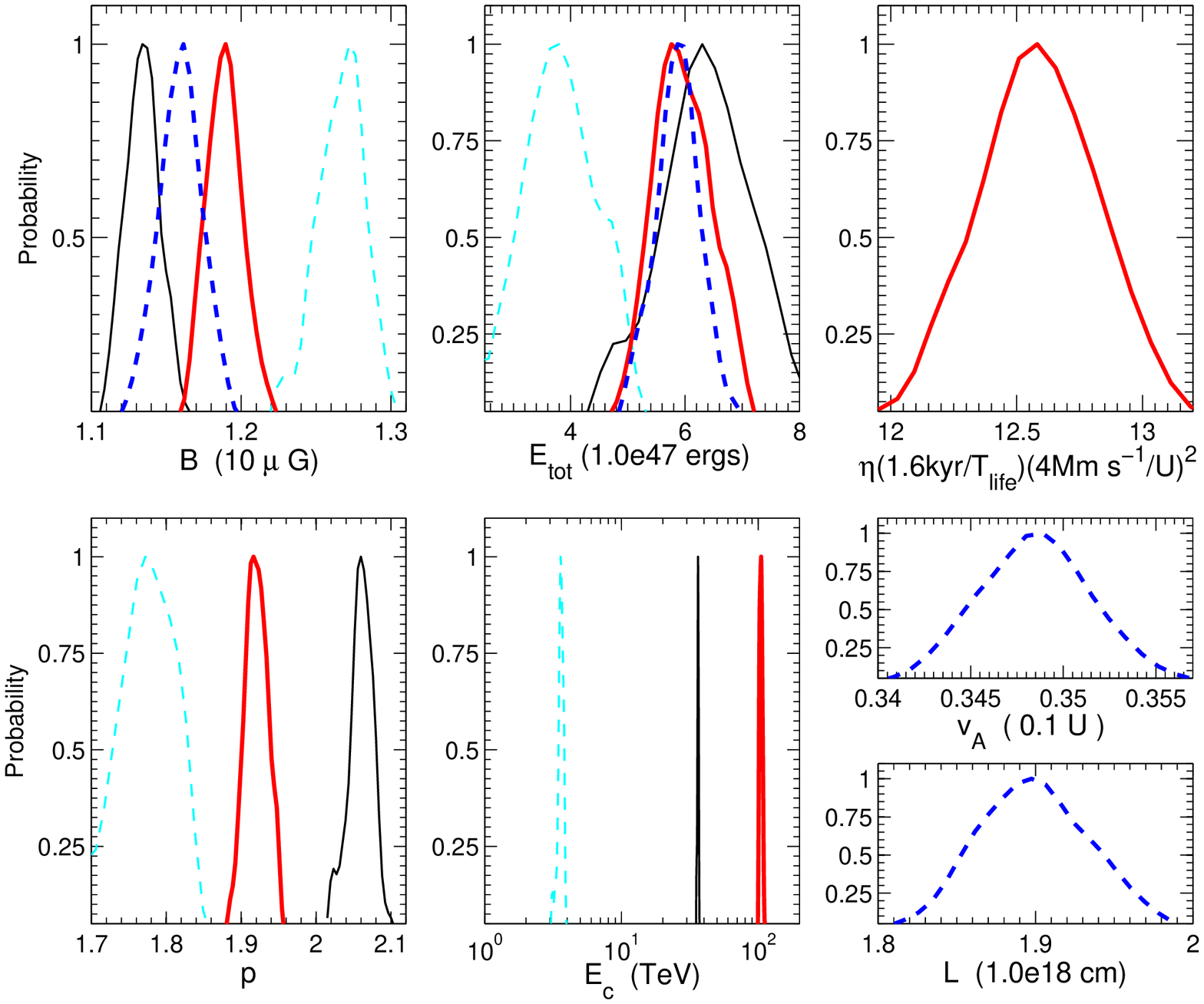}
\caption{Probability density function of model parameters normalized at the peak
value. Different line types correspond to different models as explained in Figure
\ref{spec} and the text.
\label{prob}}
\end{figure}

{\bf I:} Without detailed modeling of the particle acceleration processes, a
power-law electron distribution with an exponential high-energy cutoff is often
assumed to fit the observed spectrum:
\begin{equation}
N(\gamma)\propto \gamma^{-p}\exp({-\gamma/\gamma_c})
\end{equation}
where $p$ is the spectral index, $\gamma_c$ the Lorentz factor of
the high-energy cutoff $E_c=\gamma_c m_ec^2$, $m_e$ and $c$ are the
electron mass and the speed of light, respectively. The integration
of $N(\gamma)$ over the Lorentz factor $\gamma$ gives the particle
number density. It has been shown that this model gives a
sufficient fit to previous observations of SNR RX J1713.7-3946 when
the interstellar radiation field is considered \citep{p06}; however,
more detailed studies suggest that such a model systematically
underestimates the photon flux at the low-energy end of HESS
observations \citep{t08, za10}. In this paper, we adopt the
interstellar radiation field spectrum of \citet{p06}. There are
then four model parameters. Besides the magnetic field $B$,
$p$, and $\gamma_c$, we need a normalization for the relativistic
electrons $E_{\rm {tot}}$, which is the total energy of electrons
with $\gamma>10$. Figures \ref{spec} and
\ref{prob} give the best fit of this model to the data
and the probability density of the model parameters.
All four model parameters are well constrained. However, detailed
examination of residuals of the spectral fit shows that there are
several data points below 1 and near 10 TeV with normalized
residuals significantly greater than 3.

{\bf II:} \citet{l08} suggested that the fit to the $\gamma$-ray
spectrum can be improved by considering a more gradual cutoff in
the electron distribution:
\begin{equation}
N(\gamma)\propto \gamma^{-p}\exp{-(\gamma/\gamma_c)^{1/2}}\,.
\end{equation}
Several mechanisms can cause variations in the shape of the
high-energy cutoff \citep{b88, bld06, za07, b10}. However, given the
sharp cutoff of the X-ray spectrum, \citet{t08} claim that the
electron distribution must have a very steep cutoff. Indeed, by using
their data and performing the MCMC fit, we obtain radio fluxes
nearly one order of magnitude below the observed values. The model
also significantly overestimates the hard X-ray fluxes.
To increase the weight of the radio and
$\gamma$-ray data in the model fit, we artificially increase the
errors of the X-ray data from {\it Suzaku} by a factor of 2
\footnote{Given the complexity of the instrumental background and
calibration, and unknowns, such as the estimated cosmic X-ray
background, such an artificial increase may well be reasonable.}
(Figs. \ref{spec} and $\ref{prob}$). The normalized residuals are indeed improved
significantly with a reduced $\chi^2$ of 1.2, in agreement with
Liu et al. (2008), where cruder X-ray data from {\it ASCA}
observations were used. However, as emphasized by \citet{t08},
the sharp cutoff in the X-ray spectrum is significant and the error
estimate of the {\it Suzaku} observation appears to be robust
\citep{tak08}. This simple model therefore cannot fit the X-ray spectrum.

{\bf III:} The particle distribution is determined by the injection
process and the spatial diffusion coefficient $\chi$ in the DSA model.
The time-dependent electron distribution with a constant injection
rate may be approximated as \citep{fd83, b88, d91}
\begin{equation}
N(\gamma)\propto \gamma^{-p} \exp
[-9\chi (\gamma)/U^2{\rm T_{\rm life}}]\,, \label{f2}
\end{equation}
where $T_{\rm life}$ is the supernova lifetime since the onset of
particle acceleration and $U$ the shock speed.
A very sharp high-energy cutoff can be produced when gyro-radii of
high-energy particles exceed the characteristic length of the
magnetic field in the background plasma $l_d$. Particles at such
high energies do not perform gyro-motions around the chaotic
magnetic field and instead interact randomly with the background
fields. The spatial diffusion coefficient of these particles
increases as the square of the relativistic Lorentz factor in the
absence of waves on scales greater than $l_d$. Consequently, they
can readily cross the acceleration site without interactions.
We assume a spatial diffusion coefficient
\begin{equation}
\chi(\gamma) = \eta \left\{
\begin{array}{ll}
{(\gamma/\gamma_c)^{\beta} \gamma_c m_ec^3/(3eB)}\ \ \ \ \ &{\rm
 for}\;\;\;\; \gamma \leq \gamma_c \,, \\
{\gamma^2 m_ec^3/(3eB\gamma_c)}\ \ \ \ \ &{\rm for}\;\;\;\; \gamma \geq
 \gamma_c \,,\\
\end{array}
\right.
\label{dsaxi}
\end{equation}
where $e$ is the elementary charge, $\eta>1$ depends on the turbulence
intensity, $\beta=1/2$ and 1/3 for particle interactions with a spectrum
of plasma waves following the \citet{k65} and \citet{k41} phenomenology
for the turbulence spectral evolution, respectively. The characteristic
length of the magnetic field is then given by the gyro-radius of
electrons at $\gamma_c$: $l_d= \gamma_c m_e c^2/e B$.

Compared with the previous two models, an extra parameter is introduced
here: $\eta/T_{\rm life}U^2$, which is proportional to the electron
acceleration timescale.
Figures \ref{spec} and \ref{prob} show the results with $\beta=1/2$.
The probability density of
$\eta(1.6 {\rm kyr}/T_{\rm life})(4 {\rm Mm\ s}^{-1}/U)^2$
peaks near a typical value of 12.
Since $\eta$ has to be greater than 1, we have $T_{\rm
life}U^2>0.02 c$ lyr.
The gyro-radius of electrons at the
cutoff Lorentz factor $\gamma_c\sim 2\times 10^8$
is $\sim$0.01 pc comparable to the width of the observed highly variable X-ray
filaments. 
The X-ray variability may then be attributed to spatial diffusion
of X-ray emitting electrons from a high-density region formed presumably via an
intermittent process. It is interesting to note that the best fit
spectral index $p=1.92$ is less than $2$, implying acceleration by
multiple shocks or strong nonlinear effects \citep{za10}. The model with
$\beta=1/3$ gives a similar fit with a slightly higher value of $\eta$ and
sharper hard X-ray cutoff.

{\bf IV:} \citet{f10} have carried out detailed modeling of electron acceleration by
compressive motions in the shock downstream and argued that it might
also explain these observations.
Here we follow the same treatment
and also consider the effect of incompressive motions, which can dominate
the spatial diffusion and therefore the escape of energetic electrons from
the acceleration region.
The acceleration rate is given by \citep{bt93}
\begin{equation}
\tau_{ac}^{-1}= D_{\gamma\gamma}/\gamma^2 =  {4\pi \chi\over
 9}\int_{k_m}^{k_d}dk {k^4
 S(k)\over v_F^2+ \chi^2k^2}\,,
\label{tac}
\end{equation}
where $\chi$, $k$, $v_F$, and $S$ are the spatial diffusion
coefficient, the wave-number, speed, and intensity of compressive
waves, respectively. The overall wave intensity  $\int S(k) 4\pi
k^2dk= \xi v_F^2$ with $\xi<1$ for subsonic turbulence.
\footnote{Here we adopt the formula given by \citet{bt93},
which gives an acceleration rate a factor of 2 lower than that of \citet{p88}.}
The turbulence is generated on a scale $L$ near the shock front,
and $l_d$ corresponds to the scale where the turbulence speed is comparable to
the Alfv\'{e}n speed $v_{\rm A}$. Relativistic electrons with a gyro-radius less
than $l_d$ have a mean-free path comparable to $l_d$ since waves on
smaller scales are subject to strong transit time damping by
thermal background ions. We then have the diffusion coefficient
\begin{equation}
\chi(\gamma,x) = \left\{
\begin{array}{ll}
{cl_d(x)/3}\ \ \ \ \ &{\rm for}\;\;\;\; \gamma \leq \gamma_c(x) \,, \\
{\gamma m_ec^3/(3 eB)}\ \ \ \ \ &{\rm for}\;\;\;\; \gamma \geq
 \gamma_c(x) \,,\\
\end{array}
\right. \label{saxi}
\end{equation}
where $x$ indicates the spatial location in the downstream, $l_d = \gamma_c m_ec^2/eB$,
and we have assumed Bohm diffusion for $\gamma>\gamma_c$.  Due to gradual
decay of the turbulence intensity, related quantities evolve with
$x$ in the downstream. Equation (\ref{saxi}) is quite different from
equation (\ref{dsaxi}), where particle scattering by the background
turbulence is parameterized without considering details of the
turbulence structure. Considering the presence of strong turbulence
with $u>v_{\rm A}$ between the scales of $l_d$ and $L$, Bohm
diffusion may be appropriate, and as we show below, faster
diffusion may not be able to accelerate electrons efficiently to
account for the observations. The intensity of the incompressive
motions $T(k)$ can be determined from the energy conservation:
$\int [2T(k)+S(k)]4\pi k^2dk=3u^2$, where $3u^2/2$ is the total
turbulence energy density per unit mass. Further downstream, where
$v_F$ is high ($\ge 6^{1/2} u$), the incompressive motions vanish completely.
To derive the escape time $T_{esc}$ of electrons from the acceleration region,
we need to consider the enhancement of the spatial diffusion coefficient by the
turbulence motions. According to Eq. (4.30) in \citet{bt93}, the
effective diffusion coefficient $\chi_*$
can be derived from
\begin{eqnarray}
\chi_* &=& \chi + \frac{8\pi}{3\chi_*}\int_{k_m}^{k_d} T(k) dk + \\\nonumber
&&\frac{4\pi \chi_*}{3}\int_{k_m}^{k_d}
\left[\frac{1}{k^2\chi_*^2+v_F^2}-\frac{2(k^2\chi_*^2-v_F^2)}
{(k^2\chi_*^2+v_F^2)^2}\right]k^2 S(k) dk\,,
\end{eqnarray}
where we have ignored the decay of incompressive motions. Then we
have
\begin{equation}
T_{esc} = L^2/\chi_*.
\end{equation}

With the above treatment, we can study particle acceleration by
compressive motions all the way to the shock front, which is
different from the case studied by \citet{f10}, where only the
subsonic phase with $v_F>u$ is considered. Near the shock front, we assume
$\xi =1/2$ so that equation (\ref{tac}) is applicable. Since the compressive
wave intensity is lower than in \citet{f10}
and the acceleration timescale of TeV electrons becomes comparable to the
supernova lifetime, a time-dependent treatment is necessary.
\footnote{At the point, where $v_F=u$, the compressive wave intensity used
here is 3 times lower than in \citet{f10}.}
Both the acceleration and escape timescale
explicitly depend on the diffusion coefficient $\chi$, which is
independent of $\gamma$ for $\gamma<\gamma_c$. Assuming that
electrons are injected at the Lorentz factor
$\gamma_{inj}=10$ with a constant rate, the time-dependent results of
\citeauthor{bld06} (\citeyear{bld06}; see also \citealt{cs84, pp95})
can be used to describe the evolution of the electron distribution function at
energies below $\gamma_c$. Near the cutoff energy $\gamma_c$, the
nontrivial dependence of the acceleration and escape timescales on
the Lorentz factor demand numerical solutions.
We assume the time-dependent distribution function has a high-energy cutoff given by
$\exp[-(\tau_{ac}/T_{esc})^{1/2}]$ \citep[For details, see][]{f10}.
Figures (\ref{spec}) and (\ref{prob}) show the results for $U=4000$ km/s with the
Kraichnan phenomenology. The quality of this fit is comparable to the DSA
model. The deduced model parameters also agree with those obtained in \citet{f10}.

\section{Discussion}
\label{dis}

The SA model (Fig. \ref{spec}) predicts a higher hard
X-ray flux than the DSA model since the diffusion
coefficient of the DSA model has a stronger dependence on the electron
energies leading to a sharper high-energy cutoff. The electron distribution
in the SA model includes contributions from broad regions in the downstream,
which also makes the overall electron distribution broader.
Future observations with {\it NuSTAR} and {\it HXMT}
may be able to distinguish these two models.

In actual turbulent astrophysical shocks, both the DSA and SA mechanisms may
contribute to the electron acceleration \citep{a90, bt93}. The
distinction between the two resides in the dissipation structure \citep{b88}.
Supersonic shock fronts dominate the acceleration in the DSA model.
The SA model dominates in the subsonic phase. It is possible that
the shock downstream of SNR RX 1713.7-3946 has both a
supersonic and subsonic phase turbulence with the former closer to the shock
front. The electron acceleration is therefore a continuous process
in the turbulent downstream. In the SA model of this paper, we
intentionally remove the DSA by requiring that the compressive waves
be subsonic in the downstream. A self-consistent treatment of the turbulence
evolution in the shock downstream is needed for
particle acceleration near astrophysical shock fronts.

All the models appear to systematically underproduce $\gamma$-ray flux in the
{\it Fermi} energy band and overproduce flux near 10 TeV. The latter could be
caused by inhomogeneities of the background magnetic field, which implies lower
cutoff energy of the electron distribution and less high-energy $\gamma$-ray
flux \citep{t08}. The former requires a broader electron distribution from the
GeV to TeV energy range. A combination of DSA and SA model with the former
dominating the higher energy particle acceleration and the latter enhancing
the lower energy ones may address this issue. Such a scenario is possible
if both supersonic and subsonic turbulence are produced by the supernova
shock \citep{b88, bt93}.
Contributions to GeV--TeV emission from decay of neutral pions
produced by inelastic collisions of relativistic protons with the
background ions may also explain these residuals. However, as shown by
\citet{e10} and  \citet{kw08}, relativistic protons can not be the
dominant TeV emission component \citep{za10}. Including this
component is beyond the scope of the current investigation but may give
a good constraint on the relative acceleration efficiency of
relativistic electrons and protons \citep{kw08}.

\section{Conclusions}
\label{con}

With recent high spectral-resolution observations of SNR RX 1713.7-3946 in
X-rays and $\gamma$-rays, we show that, in general, models with simple
parametric descriptions of energetic particle distribution give a poor fit to
the broadband spectrum. Details of relativistic electron acceleration may
be probed with advanced models. Although both the DSA and SA model give
acceptable fits to the spectrum, the SA model predicts a higher hard X-ray
flux than the DSA model, which may be tested by future observations. The
spatial distribution of energetic electrons in the SA and DSA models is
also different, and the source structure can be used to distinguish these
two models as well. The results agree with the scenario where
energetic electrons are accelerated by interacting diffusively with a
turbulent electromagnetic field and produce nearly all of the observed
emissions from the shell of SNR RX 1713.7-3946.  Turbulence evolution at
astrophysical shocks appears to play a key role in advancing our
understanding of particle acceleration processes.

\begin{acknowledgements}
This work is supported in part by the SOLAIRE research and training
network at the University of Glasgow (MTRN-CT-2006-035484), the
National Science Foundation of China (grants 10963004 and 10778702),
Yunnan Provincial Science Foundation of China (grant 2008CD061) and
SRFDP of China (grant 20095301120006). SL thanks Eduard Kontar for
helpful discussion. QY thanks Jie Liu for helping to develop the
MCMC code adapted from the COSMOMC code of \citet{l02}.
\end{acknowledgements}

\bibliographystyle{aa}

\begin{thebibliography}{}
\bibitem[Achterberg (1990)]{a90}
Achterberg, A. 1990, A\&A, 231, 251

\bibitem[Aharonian et al. (2006)]{a06}
Aharonian, F. A., et al. 2006, A\&A, 449, 223

\bibitem[Aharonian et al. (2007)]{a07}
Aharonian, F. A., et al. 2007, A\&A, 464, 235

\bibitem[Becker, Le \& Dermer (2006)]{bld06}
Becker, P. A., Le, T., Demer, C. D. 2006, ApJ, 647, 539B

\bibitem[Berezhko (1996)]{b96}
Berezhko, E. G. 1996, Astroparticle Physics, 5, 367

\bibitem[Berezhko \& Krymsky (1988)]{b88}
Berezhko, E. G., \& Krymsky, G. F.
1988, Sov. Phys. Usp., 31, 27

\bibitem[Blasi (2010)]{b10}
Blasi, P. 2010, MNRAS, 402, 2807

\bibitem[Butt (2009)]{b09}
Butt, Y. M. 2009, Nature, 460, 701

\bibitem[Bykov \& Toptygin (1993)]{bt93}
Bykov, A. M., \& Toptygin, I. N. 1993, Phys. Usp., 36, 1020

\bibitem[Cassam-Chena\"{i} et al. (2004)]{c04}
Cassam-Chena\"{i}. G., Decourchelle, A., Ballet, J., et al. 2004, A\&A, 427, 199

\bibitem[Cowsik \& Sarkar (1984)]{cs84}
Cowsik, R., \& Sarkar, S. 1984, MNRAS, 207, 745

\bibitem[Drury (1983)]{d83}
Drury, L. 1983, SSRv, 36, 57

\bibitem[Drury (1991)]{d91}
Drury, L. 1991, MNRAS, 251, 340

\bibitem[Ellision et al. (2010)]{e10}
Ellison, D. C., Patnaude, D. J., Slane, P., \& Raymond, J. 2010, ApJ, 712, 287

\bibitem[Fan, Liu, \& Fryer (2010)]{f10}
Fan, Z. H., Liu, S. M., Fryer, C. L. 2010, MNRAS, doi:10.1111/j.1365-2966.2010.16767.x


\bibitem[Forman \& Drury (1983)]{fd83}
Forman, M. A., \& Drury, L. O. 1983, ICRC, 2, 267

\bibitem[Funk (2009)]{f09}
Funk, S. 2009, Fermi Symposium.

\bibitem[Gamerman (1997)]{g97}
Gamerman, D., {\it Markov Chain Monte Carlo: Stochastic Simulation
for Bayesian Inference}, Chapman and Hall, London, 1997

\bibitem[Katz \& Waxman (2008)]{kw08}
Katz, B., \& Waxman, E. 2008, JCAP, 01, 018

\bibitem[Kolmogorov (1941)]{k41} Kolmogorov A., 1941, Dokl. Akad. Nauk SSSR, 30, 301

\bibitem[Kraichnan (1965)]{k65}
Kraichnan 1965, Phys. Fluids, 8, 1385

\bibitem[Lewis \& Bridle (2002)]{l02}
Lewis, A.,  \& Bridle, S. 2002, Phys. Rev. D 66, 103511

\bibitem[Liu et al. (2008)]{l08}
Liu, S. M., Fan, Z. H., Fryer, C. L., Wang, J. M., Li, H., 2008, ApJ,
683, L163

\bibitem[MacKay (2003)]{m03}
MacKay, D., {\it Information Theory, Inference, and Learning Algorithms},
Cambridge University Press, 2003

\bibitem[Neil (1993)]{n93}
Neal, R. M., {\it Probabilistic Inference Using Markov Chain Monte Carlo
Methods}, Technical Report CRG-TR-93-1, Department of Computer Science,
University of Toronto, 1993

\bibitem[Park \& Petrosion (1995)]{pp95}
Park, B. T., \& Petrosian, V. 1995, ApJ, 446, 699

\bibitem[Plaga (2008)]{p08}
Plaga, R. 2008, NewA, 13, 73

\bibitem[Porter et al. (2006)]{p06}
Porter, T. A., Moskalenko, I. V., \& Strong, A. W. 2006, ApJ, 648,
L29

\bibitem[Ptuskin (1988)]{p88}
Ptuskin, V. S. 1988, Soviet Astron. Lett., 14, 255

\bibitem[Takahashi et al. (2008)]{tak08}
Takahashi, T., et al. 2008, PASJ, 60, 131

\bibitem[Tanaka et al. (2008)]{t08}
Tanaka, T., et al. 2008, ApJ, 685, 988

\bibitem[Toptygin (1980)]{t80}
Toptygin, I. N. 1980, SSRv, 26, 157


\bibitem[Uchiyama et al. (2007)]{u07}
Uchiyama, Y., et al. 2007, Nature, 449, 576

\bibitem[Zirakashvili \& Aharonian (2007)]{za07}
Zirakashvili, V. N., \& Aharonian, F. A. 2007, A\&A, 465, 695

\bibitem[Zirakashvili \& Aharonian (2010)]{za10}
Zirakashvili, V. N., \& Aharonian, F. A. 2010, ApJ, 708, 965

\end{thebibliography}

\end{document}